\documentclass[10pt,twocolumn,letterpaper]{article}

\usepackage{cvpr}              

\usepackage[dvipsnames]{xcolor}

\definecolor{cvprblue}{rgb}{0.21,0.49,0.74}
\usepackage[pagebackref,breaklinks,colorlinks,allcolors=cvprblue]{hyperref}
\usepackage{multirow}
\usepackage{algorithm}
\usepackage{algorithmic}
\usepackage{makecell}

\def\paperID{5136} 
\def\confName{CVPR}
\def\confYear{2025}

\title{SGCR: Spherical Gaussians for Efficient 3D Curve Reconstruction}

\author{Xinran Yang$^1$, Donghao Ji$^1$, Yuanqi Li$^1$, Jie Guo$^1$, Yanwen Guo$^{1*}$, Junyuan Xie$^1$\\
$^{1}$Nanjing University, Nanjing, China\\
{\tt\small \{xryang,donghaoji\}@smail.nju.edu.cn}\\
{\tt\small \{yuanqili,guojie,ywguo,jyxie\}@nju.edu.cn}
}

\begin{document}
\twocolumn[{
\renewcommand\twocolumn[1][]{#1}
\maketitle

\vspace{-5mm}
\begin{center}
    \captionsetup{type=figure}
    \includegraphics[width=\textwidth]{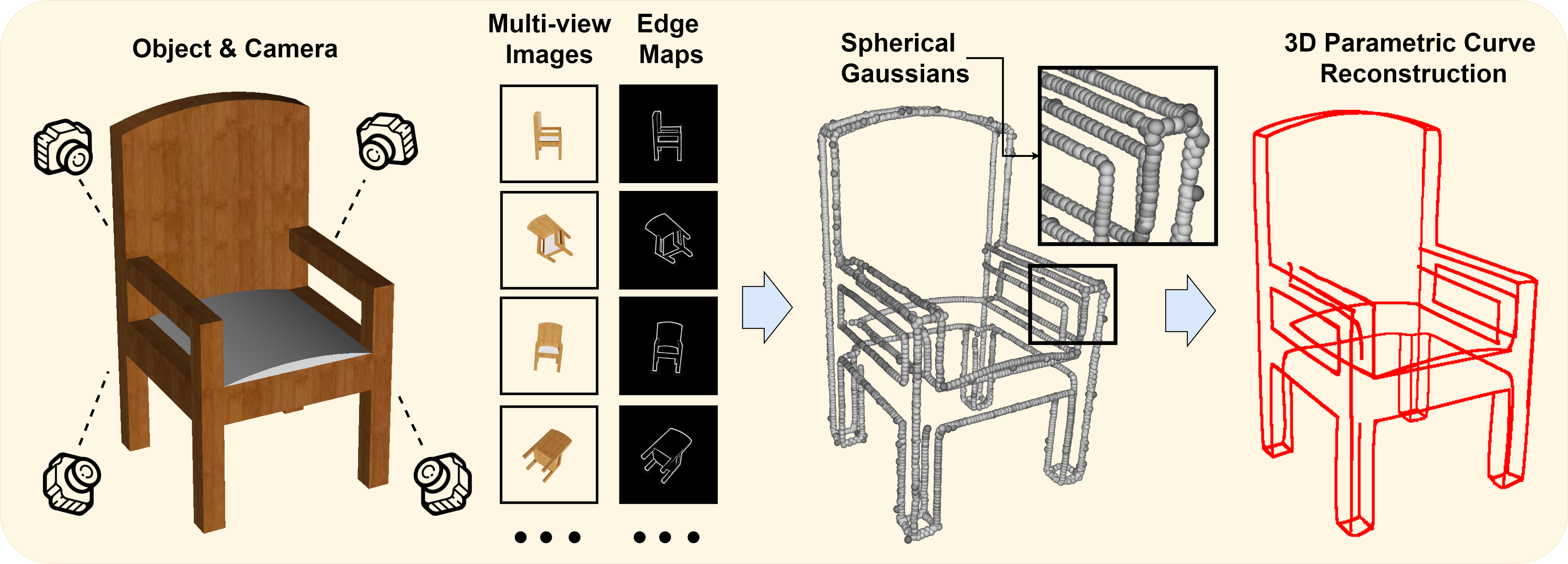}
  \caption{We present a novel method using 2D images to generate \emph{Spherical Gaussians}, enabling the accurate representation of 3D edges. Spherical Gaussians combine the advantages from 3D Gaussian Splatting and the ability to preserve excellent 3D structures. With a corresponding curve extraction algorithm, Spherical Gaussians efficiently reconstruct 3D edge curves by only 2D supervision.}
  \label{fig:teaser}
\end{center}

}]

\renewcommand{\thefootnote}{} 
\footnotetext{$^*$Corresponding author.}


\begin{abstract}
Neural rendering techniques have made substantial progress in generating photo-realistic 3D scenes. The latest 3D Gaussian Splatting technique has achieved high quality novel view synthesis as well as fast rendering speed. However, 3D Gaussians lack proficiency in defining accurate 3D geometric structures despite their explicit primitive representations. This is due to the fact that Gaussian's attributes are primarily tailored and fine-tuned for rendering diverse 2D images by their anisotropic nature. To pave the way for efficient 3D reconstruction, we present Spherical Gaussians, a simple and effective representation for 3D geometric boundaries, from which we can directly reconstruct 3D feature curves from a set of calibrated multi-view images. Spherical Gaussians is optimized from grid initialization with a view-based rendering loss, where a 2D edge map is rendered at a specific view and then compared to the ground-truth edge map extracted from the corresponding image, without the need for any 3D guidance or supervision. Given Spherical Gaussians serve as intermedia for the robust edge representation, we further introduce a novel optimization-based algorithm called SGCR to directly extract accurate parametric curves from aligned Spherical Gaussians. We demonstrate that SGCR outperforms existing state-of-the-art methods in 3D edge reconstruction while enjoying great efficiency. Code is available at \href{https://github.com/Martinyxr/SGCR}{https://github.com/Martinyxr/SGCR}.

\end{abstract}

\vspace{-4mm}

\section{Introduction}
In computer graphics and vision, photorealistic novel view synthesis (NVS) and accurate geometry reconstruction are two significant objectives. 3D Gaussian Splatting \cite{3DGS} offers an appealing explicit representation---3D Gaussians, resulting in high-quality synthesis of novel views and real-time rendering speed. However, it falls short in capturing accurate 3D geometric structures since the anisotropic 3D Gaussians, which are intended for efficient rasterization and sufficient expression ability, do not align with the geometric features (surfaces, edges and corners) in a faithful manner. A question naturally arises: Can we utilize 3D Gaussians for the direct representation of intricate geometric structures while retaining their benefits in 2D rendering?

Several recent works attempt to directly address the task of sdf generation or mesh reconstruction \cite{SuGaR, NeuSG, GSDF, 2DGS} with 3D Gaussians.
In this paper, we try to answer this question from a new perspective---reconstructing 3D parametric curves with `Spherical Gaussians'. Following the settings of classic rendering works \cite{NeRF, 3DGS}, we use a set of calibrated multi-view images as input, but our goal substitutes novel view synthesis for 3D edge/curve reconstruction. 

Feature curves serve as critical geometric cues in characterizing the structure of 3D shapes, benefiting surface reconstruction, shape perception or other downstream tasks. Traditional methods \cite{gumhold2001feature, EAR,dey1999curve,kumar2004curve} usually work with the direct representation of 3D shapes, such as polygonal meshes or point clouds. However, the presence of sharp edges may be compromised or entirely overlooked on account of imperfect 3D scanning and acquisition, especially when dealing with sparse and noisy data. In recent years, learning based-methods \cite{PIE-Net, PC2WF, DEF, NerVE} are proposed to mitigate this, but obtaining precise ground-truth labels for 3D edge supervision involves laborious manual annotation, and inevitably limiting their generality. On the contrary, 2D images is easily accessible and allow for straightforward identification of feature edges through mature edge detection methods, making 2D supervision more attractive for edge processing tasks. Despite advantages in detection and self-supervision, there still remains a huge obstacle to overcome---establishing correspondence between edges across different views when working with multi-view images, in other words, how to combine information from multiple views to construct a complete 3D structure? This is exactly the factor for our design of Spherical Gaussians (as well as SGCR), which serve as a robust bridge to connect 3D depiction with 2D guidance.

\begin{figure}[t]
  \centering
  \includegraphics[width=\linewidth]{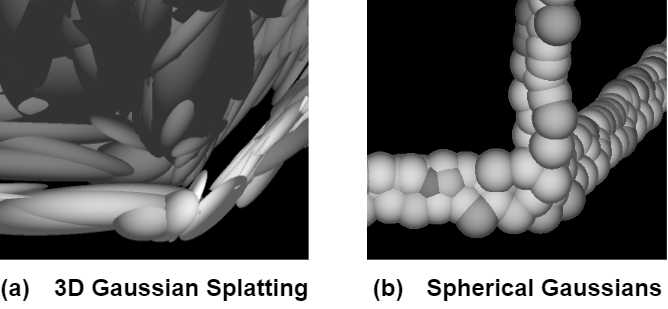}
  \caption{Visualization for optimized Gaussians. (a) Gaussians from original 3D Gaussian Splatting. (b) Our Spherical Gaussians.}
  \vspace{-5mm}
  \label{figure:Gaussian_vis}
\end{figure}

Our method consists of two main stages: Spherical Gaussians generation and parametric curves extraction. During the first stage, we utilize 2D edge maps to refine Gaussians using a view-based rendering loss compared to the ground-truth edge map derived from the image of that view. Optimizing Gaussians directly as vanilla 3D Gaussian Splatting \cite{3DGS} is problematic due to the sparse distribution of edge pixels and the diversity of Gaussian attributes. The outcome Gaussians may recover 2D images to some extent but exhibit redundant irregular ellipsoids (see Fig. \ref{figure:Gaussian_vis} (a)) and lack solid 3D geometric meanings, making following reconstruction tasks quite difficult. Therefore, we implement several constraints to transform standard 3D Gaussians into Spherical Gaussians. We introduce grid initialization, fixing-covariance training and a two-phase pruning strategy to guide the optimization process. Our final distribution of gaussians lie faithfully on sharp edges of objects in 3D space, appearing the same shape like spheres (see Fig. \ref{figure:Gaussian_vis} (b)) and holding attributes in opacity and gray-scale color.

\begin{figure}[t]
  \centering
  \includegraphics[width=\linewidth]{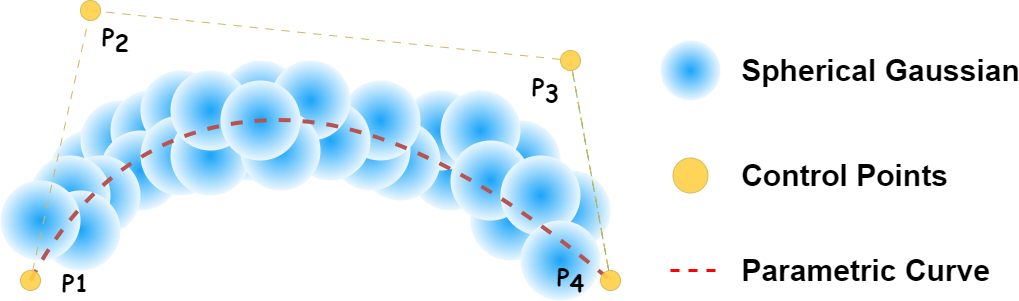}
  \caption{Illustration for fitting Bézier curve from Spherical Gaussians. $P_1$, $P_2$, $P_3$ and $P_4$ are weighted control points of the curve.}
  \vspace{-5mm}
  \label{figure:curve_vis}
\end{figure}

In the second stage, we utilize these discrete Spherical Gaussians to extract smooth 3D parametric curves for final reconstruction of edges. We present a novel optimization-based algorithm called SGCR, which involves fitting edges represented by Spherical Gaussians onto rational Bézier curves. The process begins with line fitting and then gradually refines the curves by global optimization. The final results are stored by all control points and weights in these curves. An illustration for curve extraction from Spherical Gaussians is shown in Fig. \ref{figure:curve_vis}.

We evaluate our method using a variety of object categories with complex models and scenes from ABC \cite{ABC}, ModelNet \cite{ModelNet}, DTU \cite{DTU} and Replica \cite{Replica} datasets. Extensive experiments show that SGCR, which utilizes Spherical Gaussians as bridge to reconstruct 3D parametric curves by 2D supervisions, outperforms existing state-of-the-art methods on all metrics and achieves great efficiency. In summary, our contributions include:

\begin{itemize}
\item We propose a new 3D edge extraction method from multi-view 2D images using Spherical Gaussians (SG).
\item We introduce a new optimization-based algorithm (SGCR) tailored for directly reconstructing 3D parametric curves from our generated Spherical Gaussians.
\item Our approach achieves the state-of-the-art performance as well as high efficiency on curve reconstruction tasks.
\end{itemize}

\section{Related Work}
\subsection{Image-based 3D Representation and Rendering} Image-based rendering (IBR) techniques use a collection of 2D images of a scene to generate the visual representation of the scene and render novel views. Significant advancements have been achieved by Neural Radiance Fields (NeRF) \cite{NeRF}, which depicts a scene's geometry and visual attributes as a radiance field, enabling the retrieval of color and volume density at any point in space and from any viewing angle for the purpose of rendering. Recently, the unstructured GPU-optimized 3D Gaussian Splatting (3DGS) \cite{3DGS} presents the explicit Gaussian primitive that achieves enhanced rendering speeds and improved quality without the reliance on neural components. It has also been quickly extended to multiple domains \cite{xie2023physgaussian, zielonka2023drivable, SuGaR, scaffoldgs, Yu2024MipSplatting}. We convert standard 3D Gaussians into Spherical Gaussians, enabling significant advantages in both view rendering and the representation of geometric structures.

\subsection{3D Curve Reconstruction} 
The convention of 3D curve reconstruction is based on point cloud. Traditional methods \cite{lee2000curve,amenta1998crust,dey1999curve,kumar2004curve,wang2006fitting} focus on local geometric properties of point clouds, such as normals \cite{demarsin2007detection, weber2010sharp}, curvatures \cite{yang2014automated} and hierarchical clustering \cite{feng2014fast}. Data-driven methods \cite{PC2WF,PCED-NET,DEF,wang2020reconstruction} often adopt edge detection as a binary classification for point clouds. As network architectures advance, the classifier used for edge detection has become more complex models \cite{PIE-Net, ECNet,Pointnet++, EDC-Net}. RFEPS \cite{RFEPS} presents a multistage optimization to augment the point set so that sufficient points are generated on potential edges of geometry. NerVE \cite{NerVE} presents a neural volumetric edge representation learned through a volumetric learning framework.  

Extraction curves from multi-view images was pioneered by \cite{bartoli2005structure} which introduces a full SfM system using line segments. With the developments of line detections \cite{pautrat2021sold2, pautrat2023deeplsd, xue2023holistically} and matching \cite{abdellali2021l2d2,pautrat2021sold2,pautrat2023gluestick}, several works \cite{hofer2017efficient, wei2022elsr, he2018pl, lim2022uv, LiMAP} have attempt to revisit the line mapping problem, but they are restricted to straight lines and frequently generate fragmented small lines when dealing with curves. Lifting 2D sketches/curves to 3D curves \cite{schmidt2009analytic, rivers20103d, xu2014true2form, gryaditskaya2020lifting} has also been exploited, but the ability to generate parametric curves is still limited. Recently, NEF \cite{NEF} learns a neural implicit field to capture the distribution of edge densities, and
EMAP \cite{EMAP} encodes 3D edge distance
and direction in Unsigned Distance Functions (UDF) from multi-view edge maps. However, they all require large training time for even a simple scene. In contrast, our method employs Spherical Gaussians as the direct representation for edges, achieving 3D parametric curve reconstruction both accurately and efficiently.

\begin{figure*}[h]
  \centering
  \includegraphics[width=\linewidth]{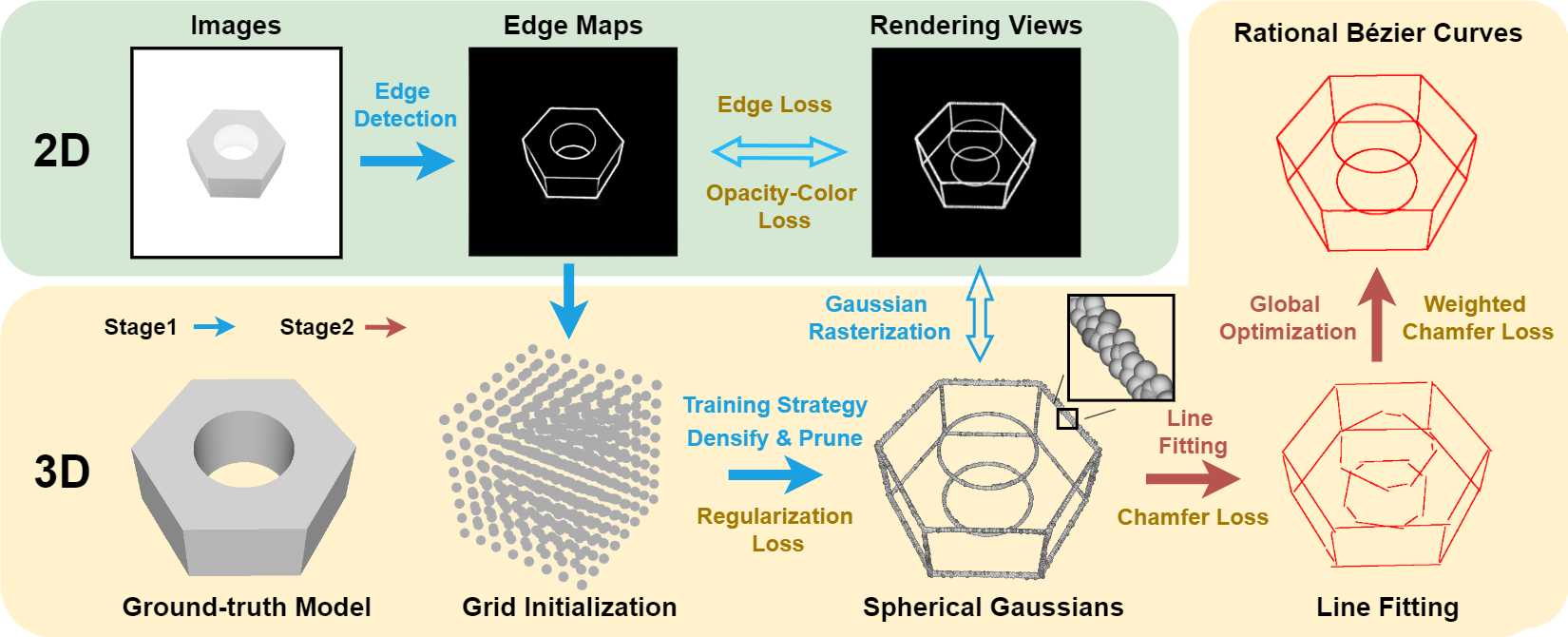}
  \caption{Pipeline illustration across 2D and 3D modality. \textbf{Stage1}: Spherical Gaussians generation. \textbf{Stage2}: Parametric curves extraction.}
  \label{fig:method_main}
  \vspace{-2mm}
\end{figure*}

\section{Method}

\subsection{Preliminary}
\noindent \textbf{3D Gaussian Splatting.} 3DGS \cite{3DGS} is a recently popular method for representing 3D scenes with 3D Gaussian primitives, achieving both explicit representation and high-quality real-time rendering. Each Gaussian primitive is defined by a covariance matrix $\mathbf{\Sigma}$, and a center point $p$ which is the mean of the Gaussian. The 3D distribution can be represented as: $ G(x) = e^{-\frac{1}{2}(x - p)^{\mathrm{T}}\mathbf{\Sigma}^{-1}(x - p)}$.

A complete 3D Gaussian is given by its mean/center $p \in \mathbb{R}^{3}$, color represented with spherical harmonics coefficients $\mathbf{H} \in \mathbb{R}^{k}$, opacity $\alpha \in \mathbb{R}$, quaternion $\mathbf{r} \in  \mathbb{R}^{4}$, and scaling factor $\mathbf{s} \in \mathbb{R}^{3}$. For a given pixel, the combined color and opacity from multiple Gaussians are weighted by Eq. 3. The color blending for overlapping points is: $ \mathbf{C} = \sum_{i \in N} \mathbf{c}_{i}\alpha_{i} \prod_{j=1}^{i-1}(1 - \alpha_{j})$ where $\mathbf{c}_{i}, \alpha_{i}$ denote the color and density of a point.

\noindent \textbf{Spherical Gaussians.} Our Spherical Gaussians enjoy the same nature of 3D Gaussians, but make the following changes for geometry priority and simplicity: (1) Remove the covariance matrix $\mathbf{\Sigma}$ of Gaussians (attributes of scaling $\mathbf{S}$ and rotation $\mathbf{R}$) and replace it with a fixed radius $r_0$, i.e. convert Gaussian primitives from various ellipsoids into spheres of uniform size. (2) Remove spherical harmonics coefficients  $\mathbf{H}$ and modify color to one-dimensional gray value (just for edge map rendering). According to our experiments, these changes do not hamper the backward propagation of gradients from 2D edge-map supervision, but impose great regularization on geometry distribution of 3D isotropic Gaussians, facilitating reconstruction tasks.

\noindent \textbf{Anisotropic Gaussians v.s. Isotropic Gaussians.} Traditional 3DGS \cite{3DGS} and its subsequent works mainly center on the technique of `splatting'---utilizing different anisotropic Gaussian splats to accurately represent the entire scene. However, our work focuses on each `Gaussian' itself, essentially displaying a reverse process: decomposing the splatting results into small `atoms' and assigning clear geometric meanings to every individual Gaussian. A thin and elongated Gaussian seems to be more efficient in representing edges, but it can be split into smaller `basis-Gaussians' with few sacrifices but for better interpretation and convenience, especially when following pointcloud-related works. That's why we favour Spherical Gaussians and more discussions can be found in suppplementary materials.

\subsection{Spherical Gaussians Generation}
\subsubsection{Initialization}
Our full pipeline is illustrated in \cref{fig:method_main}.  Given a set of calibrated multi-view images, we first adopt edge detection method \cite{PiDiNet} to generate 2D edge maps for each view. 3DGS uses structure-from-Motion (SfM) \cite{SfM} points as initialization for Gaussians, but this approach is not ideal for handling edge maps. Thus, we design grid initialization for Spherical Gaussians. We create a $50\times50\times50$ grid with points evenly distributed in space of $[0, 1]^{3}$ (see \cref{fig:method_main}---\textbf{Grid initialization}) as the initial centers for our Spherical Gaussians. This setting is better than random initialization in producing neat structures. And we choose a constant radius $r_0$ to replace the covariance matrix with Spherical Gaussians of the same size. In our experiments, we use $r_0 = 0.005$.

\subsubsection{Training Losses} 
\noindent \textbf{Edge Loss.} The original loss functions in 3DGS \cite{3DGS} to train 3D Gaussians for each view is defined as:
\begin{equation}
    \mathcal{L} = (1 - \lambda)\mathcal{L}_{1} + \lambda \mathcal{L}_{D-SSIM}
\end{equation}
where $\mathcal{L}_{1}$ and $\mathcal{L}_{D-SSIM}$ are supervised by the ground-truth view. 
However, in edge maps, edge pixels are often sparsely distributed. Relying solely on this loss function will easily make the training process stuck in local optima, causing the opacities and colors for all Gaussians to converge towards zero and resulting in all rendered images becoming black. 
Hence, we opt for a new edge-aware loss $\mathcal{L}_{edge}$ instead of $\mathcal{L}_{1}$ in order to balance the impact of both edge and non-edge pixels within each edge map. The edge-aware loss is defined as:
\begin{equation}
\mathcal{L}_{edge} = \frac{N_{I} - |E_{I}|}{N_{I}}\sum_{i \in E_{I}} \lVert I_{i} - \hat{I}_{i} \rVert^{2}_{2} + \frac{|E_{I}|}{N_{I}}\sum_{i \notin E_{I}} \lVert I_{i} - \hat{I}_{i} \rVert^{2}_{2}
\end{equation}
where $N_{I}$ is the image resolution and $E_{I}$ refers to all edge pixels in current edge map (gray scale larger than threshold $\eta=0.3$), $I_{i}$ and $\hat{I}_{i}$ refers to the pixel in Gaussian-rendering image and ground-truth edge-map image respectively. This edge-aware loss enhances the ability of Spherical Gaussians to identify edges from edge pixels.

\noindent \textbf{Opacity-Color Loss.} Another issue during edge training is about occluded edges. Some 2D edge maps may lose true edge pixels due to occlusions from these image views (see the difference between \cref{fig:method_main}--\textbf{Edge maps} and \textbf{Rendering views}), while from other views the occlusion may not happen. This inconsistency among views will confuse Spherical Gaussians during training, potentially leading to unstable optimization. During training, color attribute is directly supervised by $\mathcal{L}_{edge}$ loss, while opacity is indirectly optimized and reset at certain iterations. This difference sets them apart, and equating them impedes smooth optimization. Therefore, we introduce opacity-color loss $\mathcal{L}_{oc}$ for all Spherical Gaussians to make the value of Gaussian opacity more consistent. This opacity-color loss is defined as:
\begin{equation}
\mathcal{L}_{oc} = \sum_{i=1}^{N_{SG}} \lVert o_{i} - c_{i} \rVert^{2}_{2}
\end{equation}
where $N_{SG}$ is the number of all Spherical Gaussians during current training step, $o_{i}$ and $c_{i}$ refer to opacity and color attribute of each Spherical Gaussian respectively. $\mathcal{L}_{oc}$ helps to keep consistency between edge density and color intensity during multi-view supervision. This mechanism ensures that occluded edges---represented by Gaussians with low opacity values---are less susceptible to premature pruning during optimization, resulting in more complete 3D reconstructions. It also helps rectify edge predictions missed by 2D detectors. In our experiment, after implementing $\mathcal{L}_{oc}$, Gaussian-rendered images can depict complete edge structures with no occlusion (\cref{fig:method_main}---\textbf{Rendering Views}).

\noindent \textbf{Regularization Loss.} To restrict total number of Spherical Gaussians and accelerate convergence, we add an extra regularization term $\mathcal{L}_{regular}$
on Gaussian's opacity. The regularization loss is defines as:
\begin{equation}
\mathcal{L}_{reg} = \sum_{i=1}^{N_{SG}} log(1 + \frac{o_{i}^2}{0.5})
\end{equation}
where $N_{SG}$ is the number of current Spherical Gaussians and $o_{i}$ is each Gaussian's opacity. During the training process, this loss function works implicitly because Spherical Gaussians with opacity below a certain threshold will be pruned (as in 3DGS \cite{3DGS}), allowing for a comfortable number of total Gaussian primitives for next stage.

To summarize, our final loss function for training Spherical Gaussians is presented as :
\begin{equation}
    \mathcal{L} = (1 - \lambda_{1})\mathcal{L}_{edge} + \lambda_{1}\mathcal{L}_{D-SSIM} + \lambda_{2} \mathcal{L}_{oc} + \lambda_{3} \mathcal{L}_{reg}
\end{equation}
where the balancing parameters $\lambda_{1}$, $\lambda_{2}$, $\lambda_{3}$ are set to 0.2, 2 and 0.01 in our experiments respectively.

\begin{figure}[t]
  \centering
  \includegraphics[width=\linewidth]{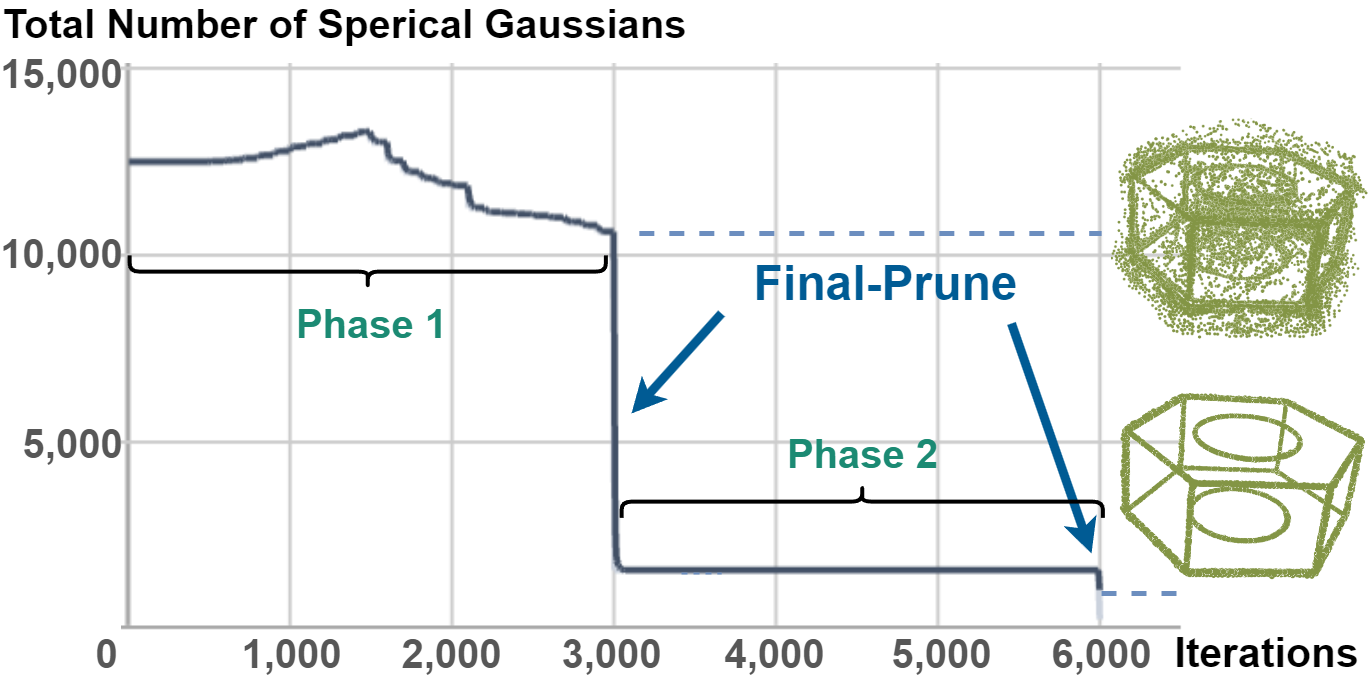}
  \vspace{-5mm}
  \caption{The change on number of Gaussians during training.}
  \label{figure:Total_Gaussians}
  \vspace{-5mm}
\end{figure}

\subsubsection{Training Strategy} 
The effective training strategy is essential for creating properly structured Spherical Gaussians. We adopt a two-phase training strategy, each phase consists of 3,000 iterations. During the first phase, we conduct densification (split and clone) of Spherical Gaussians every 200 iterations. And we reset the opacity attribute to 0.1 every 1,000 iterations. At the end of first phase, we design a \emph{final-prune}---removing all Spherical Gaussians whose opacity $o_{i} < 0.5$ and color $c_{i} < 0.1$. This operation largely reduce the total number of 
Gaussians, leading to substantial reductions in both training time and noise. During the second phase, we no longer densify Spherical Gaussians but focus on refining their positions and attributes. In the end, we perform another \emph{final-prune} to retain all significant  Spherical Gaussians (at quantity of 1k to 5k) with their attributes. 
The variation of numbers for Spherical Gaussians during all iterations is depicted in \cref{figure:Total_Gaussians}. The whole training process of Spherical Gaussians for single scene takes only about 1 minute.

\subsection{Parametric Curve Extraction from Spherical Gaussians}
After the first stage, we get the generated Spherical Gaussians (denoted as $G$) aligned with object's edges, resembling a 3D wireframe constructed by many small spheres. In the second stage (\cref{fig:method_main}---\textbf{Stage 2}) , we present of Spherical-Gaussians Curve Reconstruction (SGCR) to accomplish the final 3D edge reconstruction task. SGCR is an optimization-based algorithm guided by $G$ which includes two parts: line fitting and global optimization.  

\subsubsection{Line Fitting}

\begin{algorithm} [t]
\caption{Line Fitting}  
\label{alg:line_fitting}  
\hspace*{0.02in}{\bf Input:}
a set of Spherical Gaussians $G$ with radius $r_{0}$. \\
\hspace*{0.02in}{\bf Output:} 
the set of line endpoints $L = \{(p_i, q_i)\}_{i=1}^{|L|}$ for the coarse line-fitting of 3D edges. \\
\hspace*{0.02in}{\bf Denote:} 
$P(G)$ is the centers of $G$, and $N(0, 1)$ is the Standard Normal Distribution.

\begin{algorithmic}[1] 

\STATE $L = \emptyset$

\WHILE{$|G| > N_0$}

\STATE $S_{min}$ = MaxIntValue

\FOR{$i=1$ to $n$}  

\STATE $p, q=RandomChoice(P(G))$

\STATE $L_{p,q} = interpolate(p, q, N_s) + r_0 * N(0, 1)$

\STATE $G_{i} = \{g \in G | \underset{x \in L_{p,q}}{\min} ||P(g) - x||_{2}^{2} < \delta_{1}\} $

\STATE $p^{\prime}, q^{\prime} = \underset{p, q}{\arg\min}  \mathcal{L}_{CD}(L_{p,q}, P(G_i))$

\STATE $S_i = \underset{p, q}{\min} \mathcal{L}_{CD}(L_{p,q}, P(G_i))$  
\IF{$S_i \leq S_{min}$}

\STATE $S_{min} = S_i$, $G^{\prime} = G_i$, $(\hat{p}, \hat{q}) = (p^{\prime}, q^{\prime})$



\ENDIF
\ENDFOR

\STATE $G = G - G^{\prime}$, $L = L \cup \{(\hat{p}, \hat{q})\}$;

\ENDWHILE  

\RETURN $L$; 

\end{algorithmic}  
\end{algorithm}

We first fit a set of line segments to $G$'s centers as the coarse fitting. The pseudo code of this part is shown in \cref{alg:line_fitting}. The whole process consists of several \emph{line-fitting} turns and $G$ will be gradually fitted and partly removed by one line each turn.
\emph{Line fitting} will stop when the number of remaining $G$ is less than $N_0=5$. 

In each turn, we conduct $n$ random searches and record the best result. At the begining of $i$-th search, we randomly choose two nearest Gaussian centers as initial endpoints $p,q$ of the fitting line, and interpolate $N_s$ points along the line. To simulate the shape of Spherical Gaussians, we dilate the sampled points up by duplicating and adding Gaussian noise with standard deviation equals $G$'s radius $r_0$. Then we apply Chamfer Distance to compute the fitting loss between dilated line points and $G$. This Chamfer loss is defined as:

\begin{equation}
\begin{split}
    \mathcal{L}_{CD}(L_{p,q}, P(G_i)) & =  \frac{\gamma_1}{|L_{p,q}|} \sum_{x \in L_{p,q}} \min_{y \in P(G_i)} \lVert x - y \rVert^{2}_{2} 
    \\+\frac{1}{|P(G_i)|} & \sum_{y \in P(G_i)} \min_{x \in L_{p,q}} \lVert x - y \rVert^{2}_{2}
    \label{eq_cd}
\end{split}
\end{equation}

\begin{equation}
    G_{i} = \{g \in G | \underset{x \in L_{p,q}}{\min} ||P(g) - x||_{2}^{2} < \delta_{1}\} 
\end{equation}
where $L_{p,q}$ denotes the dilated line points between $p,q$ in $i$-th search, $P(G)$ represents Gaussian centers of all current $G$ and $\gamma_1$$=$$2$ is a balance parameter. $G_i$ is a subset of $G$ whose centers are close to $L_{p,q}$ within a threshold $\delta_{1}=0.02$. We optimize $p, q$ by minimizing \cref{eq_cd} with certain iterations and use the value of  $\mathcal{L}_{CD}$ after optimization as the final fitting score $S_{i}$ for $i$-th search.

For each turn, suppose the best fitting is in $k$-th search, i.e. $S_{k} = \min S_{i}$, we record the optimized endpoints $p^{\prime},q^{\prime}$ in $k$-th search and delete $G_k$ to update the current $G$. The next turn will then proceed accordingly. Once all the turns are completed, we will get a set of paired line endpoints $L = \{(p_i, q_i)\}_{i=1}^{|L|}$
which together form an approximate line-fitting result (\cref{fig:method_main}---\textbf{Line Fitting}).

\subsubsection{Global Optimization}

After line fitting, we resume all $G$ deleted in previous part and take opacity attribute into account as the fine fitting. We interpolate another two control points between each line endpoint pair $(p_i,q_i)$ and introduce point weight to initialize 3rd order rational Bézier curves, which is defined as:
\begin{equation}
    B(u)_{p,w}=\frac{\sum_{i=0}^{3}B_{3,i}(u)p_iw_i}{\sum_{i=0}^{3}{B_{3,i}(u)w_i}}
\end{equation}
where $B_{3,i}$ is basis functions of Bézier curves, $p_i$ is the $i$-th control points of curve and $w_i$ is the weight of control point $p_i$ ($w_i \equiv 1$ for simple Bézier curves). We opt for the rational Bézier curves over simple Bézier curves because simple Bézier curves cannot perfectly fit circles, which are frequently seen in object's edges.

We also sample $N_s$ points for each Bézier curve and dilate them up as in \emph{line fitting}. Since $G$ with higher opacity values are more significant during fine fitting, we apply the weighted Chamfer Distance to compute the fine fitting loss between global curve points and all $G$:
\begin{equation}
\begin{split}
    \mathcal{L}_{WCD}(L_{all}, P(G)) & = \frac{\gamma_2}{|L_{all}|} \sum_{x \in L_{all}} o_{y_{min}} \cdot \min_{y \in P(G)} \lVert x - y \rVert^{2}_{2} \\
    + \frac{1}{|P(G)|} & \sum_{y \in P(G)} o_{y} \cdot \min_{x \in L_{all}} \lVert x - y \rVert^{2}_{2}
    \label{eq_wcd}
\end{split}
\end{equation}
where $L_{all}$ refers to global dilated points from all curves and $P(G)$ refers to Gaussian centers of $G$, $o_y$ is the opacity attribute in $y$-th $G$ and $\gamma_2$$=$$2$ is another balance parameter. We also adopt endpoints loss to encourage connections between curves, which is defined as:
\begin{equation}
\mathcal{L}_{endpoints} = \sum_{x,y \in P_E} \min \left( \lVert x - y \rVert_2^2 - \delta_2, 0  \right) \frac{\lVert x - y \rVert_2^2}{\lVert x - y \rVert_2^2 - \delta_2}
\end{equation}
where $P_E$ is the set of endpoints (the first and last control points) and $\delta_2$ (we choose 0.01) is a threshold to only regularize endpoints which are close to each other. The final objective function to globally optimize all curves is defined as:
\begin{equation}
    \underset{\{\{z_i^j, w_i^j\}_{j=1}^4\}_{i=1}^{|L|}}{
    \arg\min} 
    \left( \mathcal{L}_{WCD} + \lambda\mathcal{L}_{endpoints}
    \right)
\end{equation}
where $z_i^j$ and $w_i^j$ are coordinates and weights of four control points in each Bézier curve, $\lambda$ is a balancing  parameter which is set to 0.005 in our experiments. The pseudo code of this part can be found in the supplementary material.

After global optimization, we finally extract 3D parametric curves from Spherical Gaussians in form of 3rd order rational Bézier curves (\cref{fig:method_main}---\textbf{Rational Bézier Curves}).

\begin{figure*}[h]
  \centering
  \includegraphics[width=\linewidth]{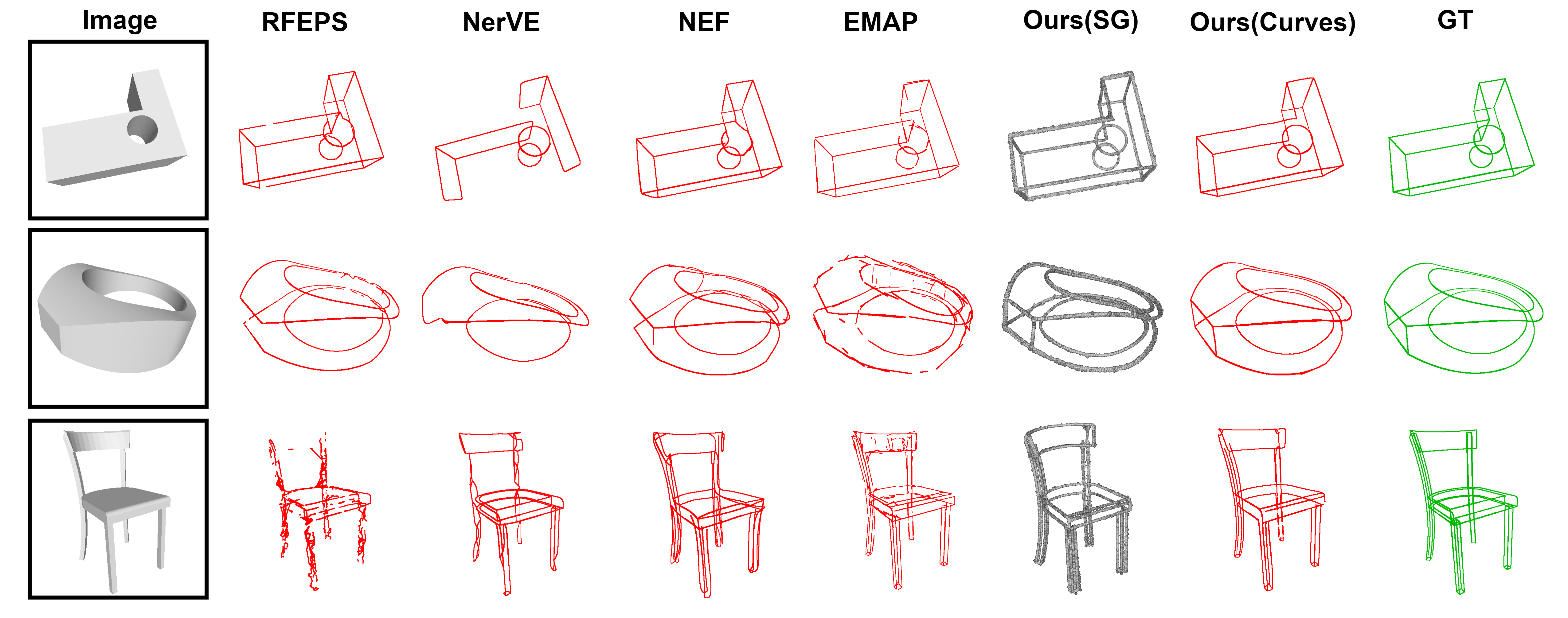}
  \vspace{-5mm}
  \caption{\textbf{Qualitative comparisons}. The first two objects come from ABC-NEF dataset and the last one comes from ModelNet dataset. \textbf{Ours(SG)} refers to our generated Spherical Gaussian in the first stage, \textbf{GT} refers to the ground-truth edge.}
  \label{figure:vis_compare1}
\end{figure*}

\setlength{\tabcolsep}{3.5pt}
\begin{table*}[h]
  \centering
  \caption{\textbf{Quantitative comparisons} on ABC-NEF Dataset \cite{NEF} and ModelNet Dataset\cite{ModelNet}. 
  }
  \resizebox{\linewidth}{!}{
  \begin{tabular}{l|ccccc|ccccc||c|cc}
    \toprule 
    \multirow{2}{*}{Method} & \multicolumn{5}{c|}{ABC-NEF Dataset} & \multicolumn{5}{|c||}{ModelNet Dataset} & Input & Training & Reconstruction\\

    \cmidrule(r){2-11}
      & CD$\downarrow$ & Precision$\uparrow$ & Recall$\uparrow$ & F-score$\uparrow$ & IoU$\uparrow$ & CD$\downarrow$ & Precision$\uparrow$ & Recall$\uparrow$ & F-score$\uparrow$ & IoU$\uparrow$ & Modality & Time $\downarrow$ & Time$\downarrow$\\
    \midrule 
    RFEPS \cite{RFEPS} & 0.0323 & 0.8962 & 0.8560 & 0.8673 & 0.8190 & 0.0731 & 0.7032 & 0.5965 & 0.6412 & 0.5621 & Point Cloud  & -- & 48s \\
    NerVE \cite{NerVE}& 0.0389 & 0.9518 & 0.7307 & 0.8267 & 0.6845 & 0.0294 & 0.9568 & 0.8017 & 0.8724 & 0.7931 & Point Cloud  & $>$10h & 1s \\
    \cmidrule{12-14}
    NEF \cite{NEF}& 0.0353 & 0.9387 & 0.8838 & 0.9044 & 0.8283 & 0.0321 & 0.9072 & 0.8158 & 0.8559 & 0.7475 & Images & $>$1h & 119s \\
    EMAP \cite{EMAP} & 0.0291 & 0.9513 & 0.8926 & 0.9210 & 0.8473 & 0.0290 & 0.9115 & \textbf{0.8488} & 0.8784 & 0.7751 & Images & $>$2h & 40s\\
    
    \cmidrule(r){2-11}
    Ours & \textbf{0.0280} & \textbf{0.9546} & \textbf{0.9052} & \textbf{0.9260} & \textbf{0.8615} & \textbf{0.0264} & \textbf{0.9627} & 0.8415 & \textbf{0.8952} & \textbf{0.8164} & Images & 87s & 32s \\

    \bottomrule 
  \end{tabular}
  }
  \label{tab: Qualitative comparison}
  \vspace{-2mm}
\end{table*}

\section{Experiments}

\subsection{Dataset}
We conduct experiments on the ABC \cite{ABC}, ModelNet \cite{ModelNet}, DTU \cite{DTU} and Replica \cite{Replica} datasets. ABC \cite{ABC} consists of massive CAD models with edge annotations, and we use a portion of it---ABC-NEF, provided by \cite{NEF}---which consisting of 115 distinct and challenging CAD models. ModelNet \cite{ModelNet} includes 3D models of everyday objects such as chairs and tables, but it does not have ground-truth edge annotations. So we manually select and annotate edges on ModelNet ground-truth meshes. Our edge dataset consists of 120 furniture models with annotated edges and is used for a more comprehensive evaluation. We utilize BlenderProc \cite{BlenderProc} to generate rendered images with these objects. We generate 50 views of $800\times800$ images for each object and adopt PiDiNet \cite{PiDiNet} to predict 2D edge maps. DTU \cite{DTU} and Replica \cite{Replica} are real-world datasets with multi-view images as input and we use it for evaluation on real-world reconstruction. 

\subsection{Comparisons}
\noindent \textbf{Settings.} We compare our method against the state-of-the-art methods in 3D curve reconstruction, including RFEPS \cite{RFEPS}, NerVE \cite{NerVE}, NEF \cite{NEF} and EMAP \cite{EMAP}. The first two requires point clouds as input, while the last two and ours use only 2D images in the same input frames and resolution. 
We all adopt their pre-trained models and corresponding reconstruction approaches to extract curves for evaluation. More experiments, discussions and details can be found in the supplementary material.

\noindent \textbf{Quantitative Comparison.} We first normalize and align all 
reconstructed curves and 
ground truth edges into the range of $[0, 1]$. Then, we sample points on all reconstructed curves and measure the distances between these points and ground-truth edge points. When assessing IoU, precision, recall, and F-score, points are considered matched if there is at least one ground-truth edge point with an $L_2$ distance smaller than 0.02. We also adopt the Chamfer Distance (CD) between point clouds to measure the geometric accuracy of reconstructed parametric curves. As reported in Table \ref{tab: Qualitative comparison}, our method outperforms other methods in all metrics on ABC-NEF \cite{NEF} and most metrics on ModelNet \cite{ModelNet}.

\noindent \textbf{Qualitative Comparison.} We illustrate the visual comparisons in \cref{figure:vis_compare1}. It can be observed that RFEPS \cite{RFEPS} and NerVE \cite{NerVE} may miss edges with weak curvature, leading to incomplete reconstruction. NEF \cite{NEF} and EMAP \cite{EMAP} struggle with handling thin structures and make errors in curve connection. On the contrary, our method benefits from aligned Spherical Gaussians (see \emph{Ours(SG)} in \cref{figure:vis_compare1}) and maintains a high level of robustness on 3D curve reconstruction.

\noindent \textbf{Time Comparison.} We also make comparisons on training time and average reconstruction time (see the last two columns in \cref{tab: Qualitative comparison}). NerVE \cite{NerVE} is a learning-based methods that rely on large point cloud datasets, requiring extensive training time. RFEPS \cite{RFEPS} is learning-free but takes time in point cloud optimization. NEF \cite{NEF} and EMAP \cite{EMAP} are supervised on images, but they all involve a time-consuming training process (several hours for single scene) for predicting edge densities or unsigned distance functions. Our SGCR  generates and optimizes only a small number of Spherical Gaussians (with two stages completing within minutes, 50$\times$ faster than NEF and EMAP), making 3D curve reconstruction quite efficient.

\subsection{Real-world Scene}

\begin{figure}[b]
  \centering
    \includegraphics[width=\linewidth]{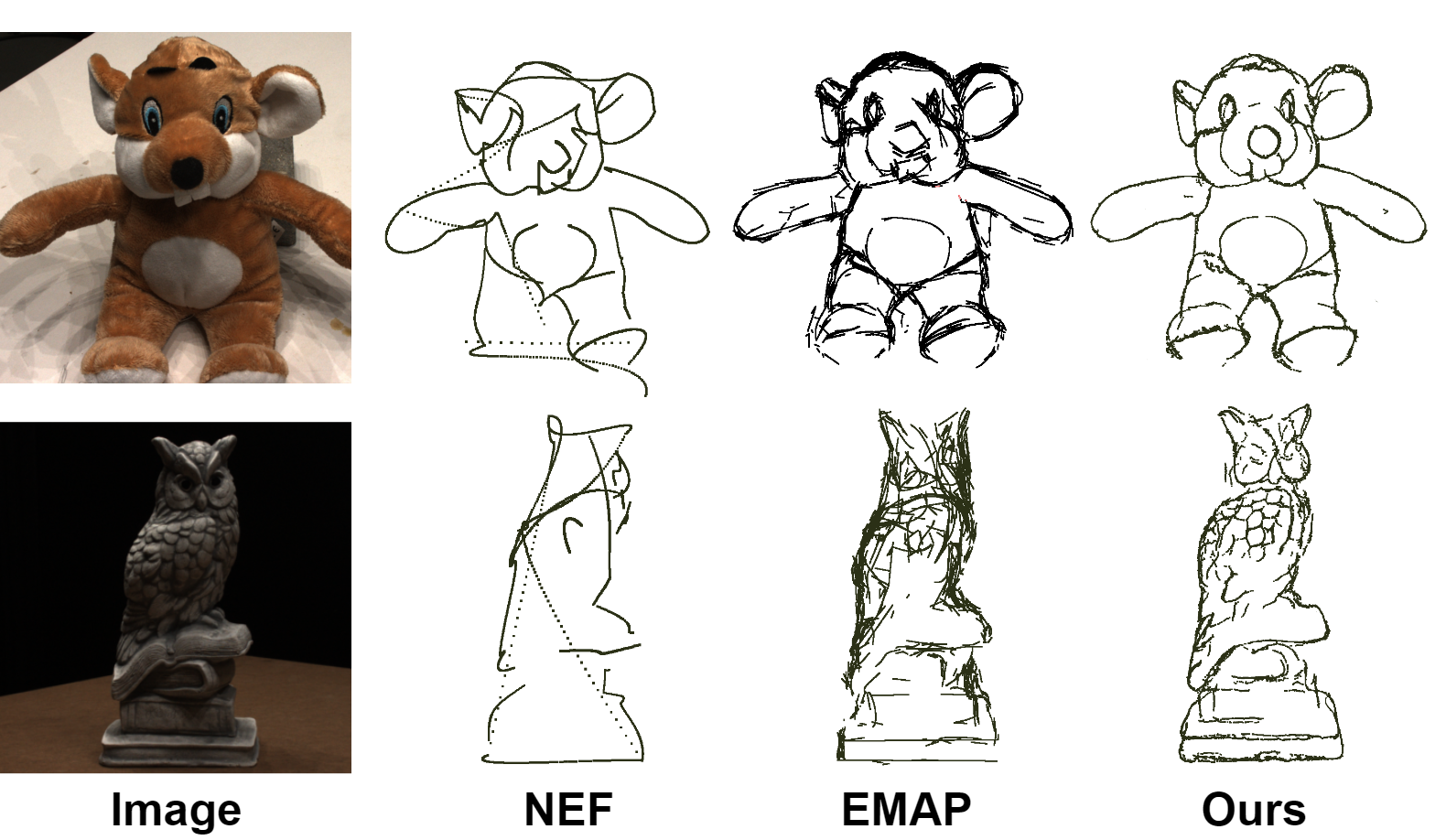}
  \caption{\textbf{Comparisons on real-world scenes from DTU \cite{DTU}.} }
  \label{figure:DTU}
\end{figure}

\begin{figure}[t]
  \centering
    \includegraphics[width=\linewidth]{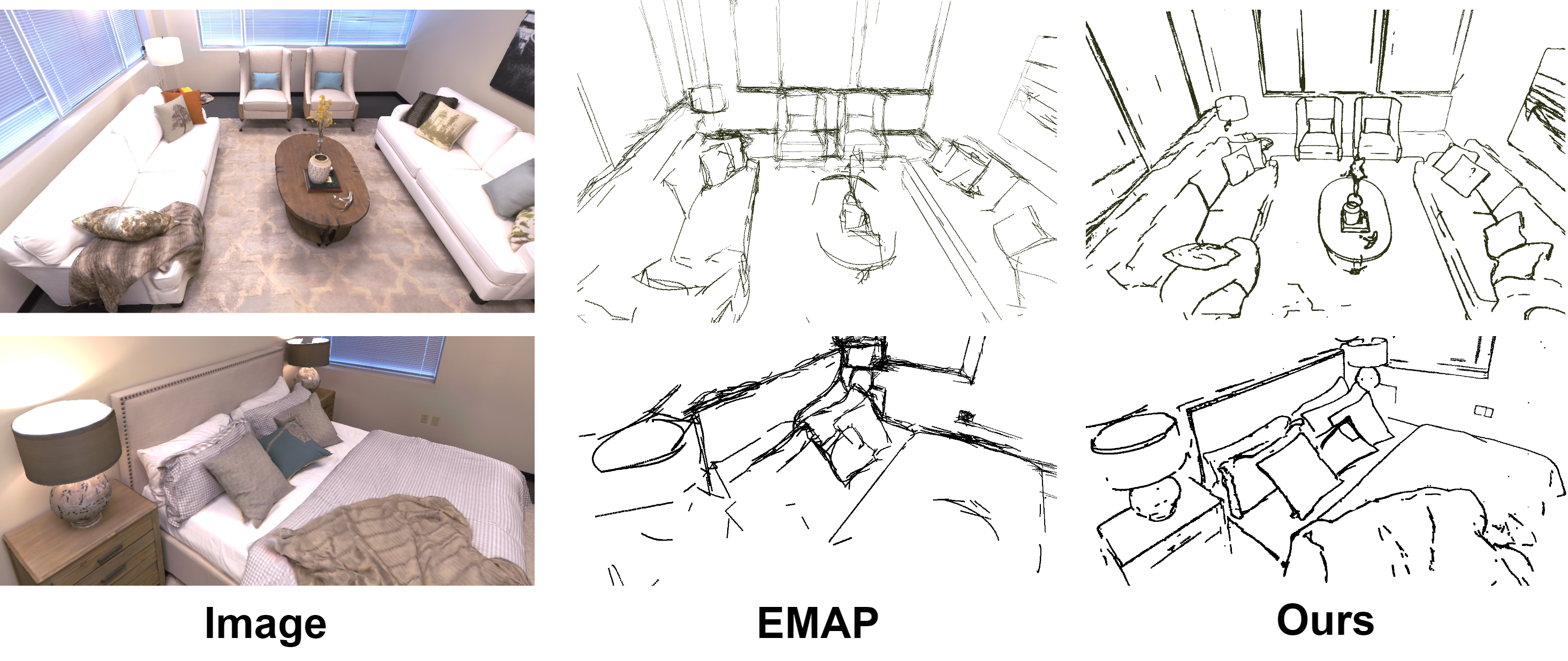}
  \vspace{-5mm}
  \caption{\textbf{Comparisons on real-world scenes from Replica \cite{Replica}.} }
  \label{figure:Replica}
  \vspace{-5mm}
\end{figure}

We further test our performance in real-world scenes on DTU \cite{DTU} and Replica \cite{Replica} dataset. They present challenging objects and complex scenes for edge reconstruction respectively. The visual comparisons against NEF \cite{NEF} and EMAP \cite{EMAP} on DTU \cite{DTU} dataset are presented in \cref{figure:DTU}. NEF misses details when fitting complex parts, and EMAP produces too many chaotic and redundant line segments, while our method reconstructs edges accurately and preserves details clearly. Another visual comparison on Replica \cite{Replica} dataset are presented in \cref{figure:Replica}. It can be seen that our method still works well on room-level edge reconstruction (NEF fails, and EMAP's training time for single real-world scene is 100$\times$ slower than ours), demonstrating our method's great generalization ability and broad applicability.

\subsection{Ablation Studies}
\noindent \textbf{Pipeline.} We conduct ablation studies on ABC-NEF dataset \cite{NEF} to evaluate the effectiveness of core designs in our method, including Sperical Gaussians, loss designs and two-phase training strategy in the first stage, as well as line fitting and global optimization in the second stage. \cref{tab:ablation} lists the quantitative results.  \emph{w/o Spherical Gaussians} stands for using original 3D Gaussians \cite{3DGS} trained by edge maps to directly represent edges and this result is quite poor, highlighting the significance of Spherical Gaussians. Statistics in \cref{tab:ablation} reveal that removing any component from our pipeline negatively impacts the final result.

 \begin{table}[h]
  \centering
  \caption{Quantitative ablation studies on our pipeline.}
  \begin{tabular}{l|ccc}
    \toprule
     Pipeline & CD$\downarrow$ & F-score$\uparrow$ & IoU$\uparrow$\\
    \midrule
    \emph{w/o Sperical Gaussians} & 0.142 & 0.330 & 0.194 \\
    \midrule
    \emph{w/o grid initialization}  & 0.032 & 0.901 & 0.822\\
    \emph{w/o edge loss $\mathcal{L}_{edge}$} & 0.052 & 0.853 & 0.714\\
    \emph{w/o opacity-color loss $\mathcal{L}_{oc}$} & 0.051 & 0.859 & 0.723\\
    \emph{w/o regularization loss $\mathcal{L}_{regular}$} & 0.029 & 0.913 & 0.846\\
    \emph{w/o two-phase training strategy} & 0.030 &  0.920 & 0.860\\
    \midrule
    \emph{w/o line fitting} & 0.030
    & 0.911 & 0.822 \\
    \emph{w/o global optimization} & 0.032
    & 0.876 & 0.757 \\
    \midrule
    Full Model & \textbf{0.028} & \textbf{0.926} & \textbf{0.862}\\
    \bottomrule
  \end{tabular}
  \label{tab:ablation}
\end{table}

\begin{table}[t]
  \centering
  \caption{Comparison on different radius choices for Spherical Gaussians. $N_{SG}$: Number of Spherical Gaussians. T1: Time for Gaussian generation. T2: Time for curve reconstruction.}
  \begin{tabular}{c|ccc||ccc}
    \toprule
     Radius $r_0$& CD$\downarrow$ & F-score$\uparrow$ & IoU $\uparrow$ & $N_{SG}$ & T1 & T2\\
    \midrule

    0.002 & 0.032 & 0.906 & 0.849 & 5,124 & 84s & 56s\\

    0.005  & \textbf{0.028} & \textbf{0.926} & \textbf{0.862} & 2,361 & 90s & 36s\\

    0.01 & 0.035 & 0.881 & 0.810 & 592 & 101s & 30s\\

    $Variable$ & 0.040 & 0.863 & 0.785 & 3,076 & 110s & 44s \\

    \bottomrule
  \end{tabular}
  \label{tab:radius}
\end{table}

\begin{figure}[h]
  \centering
  \includegraphics[width=\linewidth]{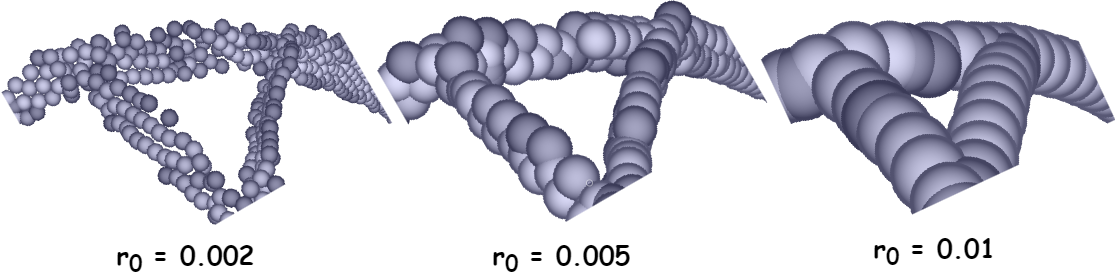}
  \vspace{-6mm}
  \caption{Visualization for Gaussians with different radii.}
  \label{figure:r0_compare}
\end{figure}

\noindent \textbf{Radius.} We analyze the effect of radius in Spherical Gaussians. We test $r_0$ = 0.002, 0.005, 0.01 and variable. Experimental results are presented in \cref{tab:radius} (setting $r_0$ as a variable does not perform well). As the radius increases, the total number of Spherical Gaussians decreases, with more time spent in the first stage and less time in the second stage. The visual comparisons can be found in \cref{figure:r0_compare}. An inappropriate radius size may lead to noise or underfitting. Using $r_0 = 0.005$ leads to the optimal reconstruction metrics as well as a reasonable running time.

\begin{figure}[t]
  \centering
  \includegraphics[width=\linewidth]{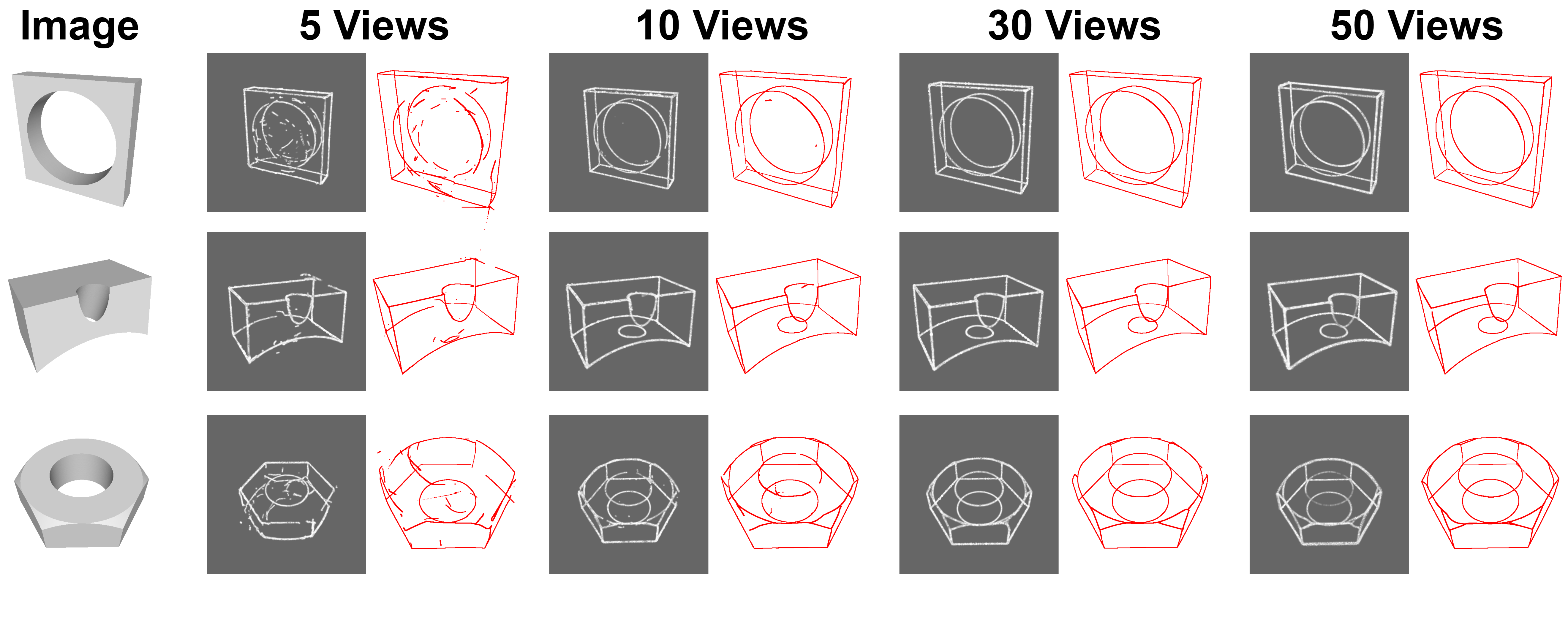}
  \vspace{-9mm}
  \caption{From left to right, we present 2D images, rendered Spherical Gaussians and final curves in 5, 10, 30 and 50 views.  }
  \label{figure:view_comparisons}
  \vspace{-3mm}
\end{figure}

\noindent \textbf{Views.} We also study the effect of the number of input views. We train Spherical Gaussians with 5, 10, 30 and 50 views respectively (all evenly distributed). The visual results are illustrated in \cref{figure:view_comparisons}. For simple cases, results with 10 views are close to satisfactory, while for complex cases, more views are required for better performance.

\section{Conclusions}
We have presented SGCR---a novel method using Spherical Gaussians for efficient 3D parametric curve reconstruction from multi-view images. The main concept involves generating Spherical Gaussians by 2D supervision and extract 3D parametric curves from them. Our method outperforms previous methods quantitatively and qualitatively, while enjoying great efficiency. SGCR also demonstrates the potential in integrating multi-modal information efficiently, bridging the gap between 2D rendering and 3D reconstruction.

\section*{Acknowledgments}
This work was supported by the National Natural Science Foundation of China under Grant 62032011.


{
    \small
    \bibliographystyle{ieeenat_fullname}
    \bibliography{main}

}


\end{document}